%% file: emnlp2023.tex
\pdfoutput=1

\documentclass[11pt]{article}

\usepackage[final]{EMNLP2023}

\usepackage{times}
\usepackage{latexsym}

\usepackage[T1]{fontenc}

\usepackage[utf8]{inputenc}

\usepackage{microtype}

\usepackage{inconsolata}

\usepackage[colorinlistoftodos,textsize=tiny]{todonotes}
\usepackage{hyperref}
\usepackage{graphicx}
\usepackage{amsmath} 
\usepackage{xspace}
\usepackage{multirow}
\usepackage{booktabs}
\usepackage{enumitem}

\title{MuRAR: A Simple and Effective Multimodal Retrieval and Answer Refinement Framework for Multimodal Question Answering}

\author{
  \textbf{Zhengyuan Zhu}\textsuperscript{1}, Daniel Lee\textsuperscript{2}, Hong Zhang\textsuperscript{2},\\
  \textbf{Sai Sree Harsha}\textsuperscript{2}, \textbf{Loic Feujio}\textsuperscript{2}, \textbf{Akash Maharaj}\textsuperscript{2}, \textbf{Yunyao Li}\textsuperscript{2}\\
  \textsuperscript{1}University of Texas at Arlington, \textsuperscript{2}Adobe\\
  \texttt{zhengyuan.zhu@mavs.uta.edu, dlee1@adobe.com, yunyaol@adobe.com }
}

\begin{document}
\input{macro}
\maketitle

\input{01_abstract}
\input{02_introduction}
\input{03_design}
\input{04_user_interface}
\input{05_dataset}
\input{06_evaluation}
\input{07_conclusion}

\newpage

\bibliography{anthology,custom}
\bibliographystyle{acl_natbib}

\newpage
\appendix
\input{08_appendix}

\end{document}

%% file: macro.tex
\newcommand{\system}[1]{{\small \ensuremath {\mathsf{#1}}}\xspace}
\newcommand{\MuRAR}{\system{MuRAR}}

%% file: 01_abstract.tex
\begin{abstract}
Recent advancements in retrieval-augmented generation have demonstrated impressive performance on the question-answering task. However, most previous work predominantly focuses on text-based answers. Although some studies have explored multimodal data, they still fall short in generating comprehensive multimodal answers, especially step-by-step tutorials for accomplishing specific goals. This capability is especially valuable in application scenarios such as enterprise chatbots, customer service systems, and educational platforms.
In this paper, we propose a simple and effective framework, \MuRAR (\textbf{Mu}ltimodal \textbf{R}etrieval and \textbf{A}nswer \textbf{R}efinement). 
\MuRAR starts by generating an initial text answer based on the user's question. It then retrieves multimodal data relevant to the snippets of the initial text answer. By leveraging the retrieved multimodal data and contextual features, \MuRAR refines the initial text answer to create a more comprehensive and informative response.
This highly adaptable framework can be easily integrated into an enterprise chatbot to produce multimodal answers with minimal modifications.
Human evaluations demonstrate that the multimodal answers generated by \MuRAR are significantly more useful and readable than plain text responses.
A video demo of \MuRAR is available at \href{https://youtu.be/ykGRtyVVQpU}{https://youtu.be/ykGRtyVVQpU}.
\end{abstract}

%% file: 02_introduction.tex
\section{Introduction}
\label{sec:introduction}

The emergence of retrieval-augmented generation (RAG) techniques~\cite{DBLP:conf/nips/LewisPPPKGKLYR020, DBLP:journals/corr/abs-2312-10997} and large language models (LLMs), such as GPT models~\cite{DBLP:conf/nips/BrownMRSKDNSSAA20, DBLP:journals/corr/abs-2303-08774}, Gemini~\cite{DBLP:journals/corr/abs-2312-11805}, and Llama~\cite{DBLP:journals/corr/abs-2302-13971}, has significantly transformed the field of question answering (QA) and improved the quality of responses generated by AI assistants.
However, the current generation of AI assistants has limitations in delivering comprehensive multimodal answers to user questions, especially step-by-step tutorials on accomplishing specific goals. 
This capability is particularly valuable in enterprise scenarios, where critical information can often be extracted from product documentation that includes multimodal data.
In such cases, images, tables, and videos are often \textit{crucial} for understanding complex, domain-specific topics. Enhancing AI assistants with the ability to incorporate multimodal information can, therefore, significantly improve user comprehension and engagement~\cite{DBLP:journals/corr/abs-2406-15000, DBLP:conf/naacl/SinghNMALS21}. This improvement offers several benefits, including increased productivity, reduced barriers to entry, higher product adoption rates, enhanced creativity, and improved user experiences.

Previous work~\cite{DBLP:conf/iclr/TalmorYCLWAIHB21, DBLP:journals/corr/abs-2004-12238, joshi2024robust} has primarily focused on leveraging various techniques to better understand multimodal data as input and generate plain text answers to a given query. In another scenario~\cite{DBLP:conf/naacl/SinghNMALS21, DBLP:journals/corr/abs-2309-05519}, the output may consist of either a text answer or a text answer accompanied by a retrieved image or video appended at the end. However, the existing solutions fail to adequately address the challenges posed by complex questions that require illustrating multiple steps to achieve a goal and integrating various multimodal content within the answer.

In summary, the main challenges are: \textbf{a)} How to retrieve the relevant multimodal data that are related and helpful to answer the user questions, and \textbf{b)} How to generate a coherent multimodal answer that integrates the retrieved multimodal data.
To address these challenges, we present \MuRAR (\textbf{Mu}ltimodal \textbf{R}etrieval and \textbf{A}nswer \textbf{R}efinement). This simple and effective framework generates coherent multimodal answers containing retrieved multimodal data, such as images, videos, and tables.
Our framework comprises three main components: text answer generation, source-based multimodal retrieval, and multimodal answer refinement. The text answer generation component retrieves relevant text documents based on the user's query and generates an initial text answer using an LLM. The source-based multimodal retrieval component retrieves multimodal data that are relevant to the text answer snippets in the initial text answer. Finally, the multimodal answer refinement component prompts an LLM to generate the final answer by integrating the retrieved multimodal data with the initial text answer.

We implemented and applied this framework on an enterprise-level AI assistant.
To verify the framework's effectiveness, we evaluated the quality of the multimodal answers on a human-annotated dataset of 300 questions and answers. 
The human evaluation results show that the \MuRAR framework effectively retrieves useful and relevant multimodal data while maintaining answer readability. Furthermore, the multimodal answers were consistently preferred over plain text answers.
Our framework can be adapted to other enterprise-level AI assistants by collecting topic-specific multimodal data and fine-tuning a topic-specific text retrieval model.
This work is, to our knowledge, the first to address the problem of generating coherent multimodal answers to user questions.

%% file: 03_design.tex
\section{Design of MuRAR}
\label{sec:methodology}

\begin{figure*}[htbp]
    \centering
    \includegraphics[width=1\textwidth]{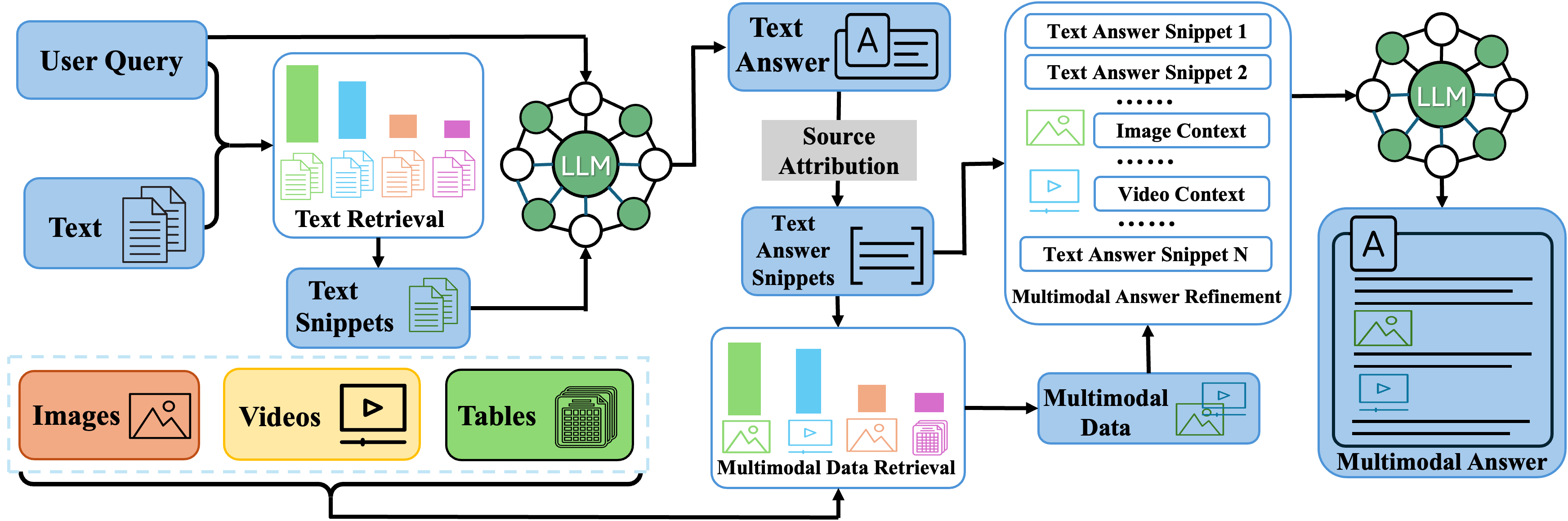}
    \caption{The architecture of the \MuRAR framework.}
    \label{fig:architecture}
\end{figure*}

\begin{figure*}[h]
    \centering
    \includegraphics[width=1\textwidth]{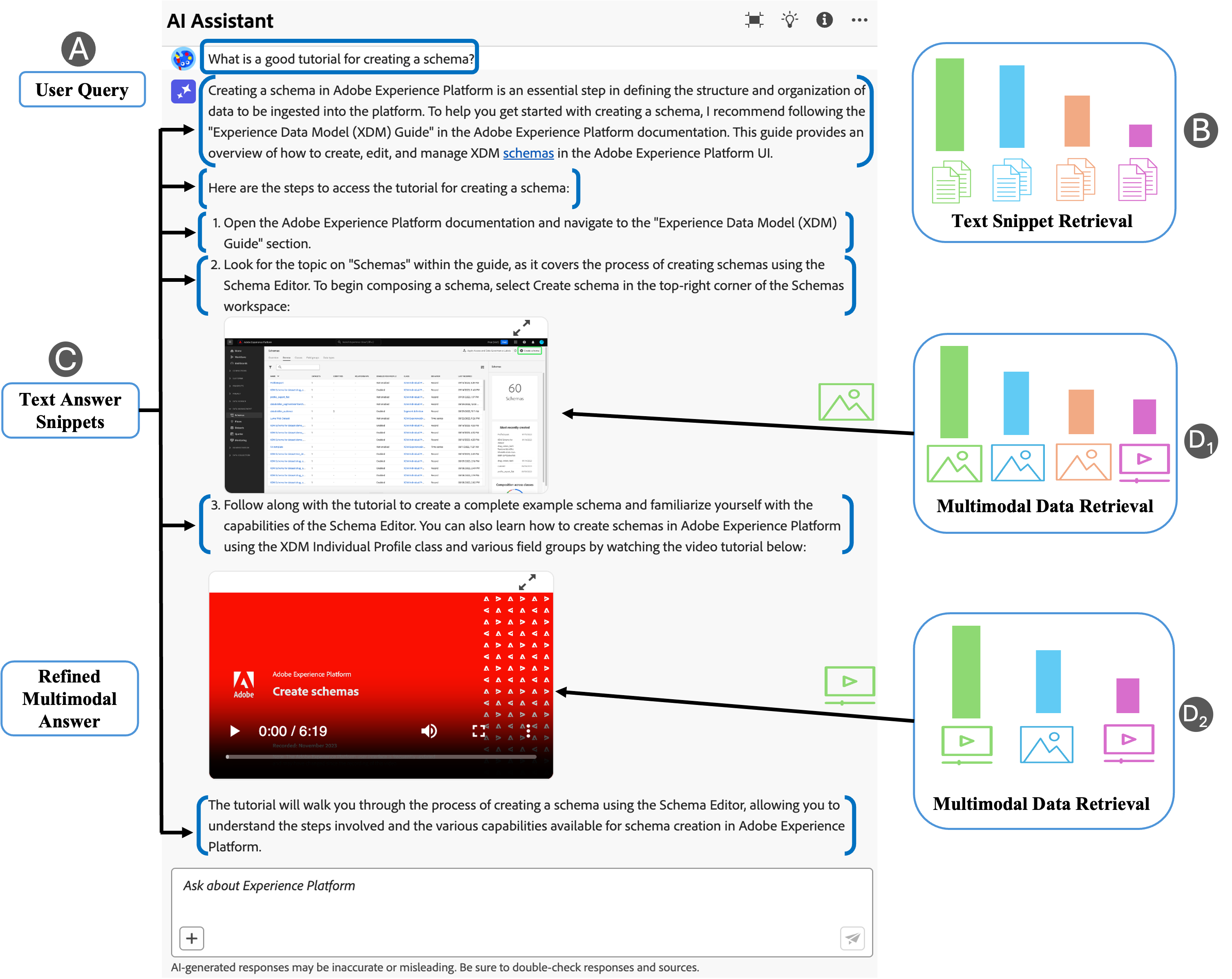}
    \caption{The interface of AI Assistant demonstrating multimodal answers is constructed by combining multimodal data retrieval and answer refinement.}
    \label{fig:UI}
\end{figure*}

Formally, given a user question $q$ as input and a set of multimodal data $\mathcal{D}=\{D_{S}, D_{I}, D_{T}, D_{V}\}$, where $\{D_{S}, D_{I}, D_{T}, D_{V}\}$ denote collections of text document snippets, images, tables, and videos, respectively, the objective is to generate a multimodal answer $A_{mm}=F(S, I, T, V)$. Here, $F$ denotes a function that organizes a set of retrieved multimodal data $(S, I, T, V) \in D$, relevant to the user's question, into a coherent and informative answer.

To achieve a high quality $A_{mm}$, we propose \MuRAR, as illustrated in Figure~\ref{fig:architecture}. 
The text answer generation component uses a RAG-style approach~\cite{DBLP:conf/nips/LewisPPPKGKLYR020, DBLP:journals/corr/abs-2312-10997}, first retrieving relevant text document snippets $S=\{s_{1}, s_{2}, ..., s_{n}\} \in D_{S}$ based on user query $q$ and then generating an initial text answer $A_{t}$ by prompting an LLM.
Next, we retrieve multimodal data, namely, $I=\{i_{1}, i_{2}, ..., i_{m}\} \in D_{I}$, $T=\{t_{1}, t_{2}, ..., t_{k}\} \in D_{T}$, and $V=\{v_{1}, v_{2}, ..., v_{l}\} \in D_{V}$ that are relevant to the text answer.
Finally, the multimodal answer refinement component generates a final multimodal answer $A_{mm}$ by incorporating the retrieved multimodal data into the initial text answer.

Notably, directly prompting or using techniques such as chain-of-thought~\cite{DBLP:conf/nips/Wei0SBIXCLZ22} with LLMs for tasks involving both text and multimodal data is ineffective for two reasons. First, the complexity of the task overwhelms the LLM as it needs to determine which data to reference, decide whether to display multimodal data and figure out where to place it within the answer. Additionally, this complexity results in low-quality multimodal answers.

\subsection{Text Answer Generation}
\label{sec:text_answer_generation}
Our text answer generation component follows the RAG-style approach. Specifically, we fine-tuned a pre-trained text embedding model~\cite{reimers-2019-sentence-bert} on an in-house annotated dataset, which includes labels indicating whether a text document snippet $s_{i}$ is relevant to a user query $q$.
The fine-tuned embedding model is then applied to the text document snippets $D_S$ to create vector indexes using FAISS~\cite{johnson2019billion}. 
These vector indexes serve as a database for retrieving relevant text document snippets by calculating the cosine similarity between the user query $q$ and each text document snippet $s_{i}$. For each user query $q$, the top five relevant text snippets are selected.
An LLM is then prompted with the user query $q$ and the retrieved top five text snippets to generate initial text answer $A_{t}$. The detailed prompt is provided in Appendix~\ref{appendix:text_answer_prompt}.

\subsection{Source-Based Multimodal Retrieval}
\label{sec:multimodal_retrieval}

The source-based multimodal retrieval component involves two steps: source attribution and section-level multimodal data retrieval.

\paragraph{Source Attribution.} 
The initial text answer $A_{t}$ is segmented into multiple sentences, with each sentence representing a continuous text answer snippet $a_{[i, j]}$ spanning from the $i$-th token to the $j$-th token in $A_{t}$.
Each text answer snippet $a_{[i, j]}$ is then compared to every text document snippet in $D_{S}$ by calculating the cosine similarity between their embeddings. The text document snippet $s_{i}$ with the highest cosine similarity score is identified as the source for $a_{[i, j]}$. 
Notably, if the highest similarity score is below 0.6, no source is assigned to $a_{[i, j]}$.
These text answer snippets and their sources are the foundation for retrieving multimodal data.

\paragraph{Section-Level Multimodal Data Retrieval.}
For each text answer snippet $a_{[i, j]}$ and its corresponding source $s_{i}$, we first locate the original web document containing $s_{i}$. We then identify the section where $s_{i}$ resides and collect all multimodal data within that section.
This approach significantly reduces the search space for multimodal data and improves the precision of the retrieval results.
The multimodal data may include images, tables, and videos. 
Multimodal data are represented using contextual and LLM-generated text features. The contextual text features are derived from the paragraphs before and after the multimodal data. For the LLM-generated text features, we utilize the image captions generated by GPT-4 and the ``alt'' text attribute extracted from the raw document HTML for images. For tables and videos, we use summarized table content and summarized video transcripts generated by the LLM, respectively.

To efficiently encode these text features, text embeddings are computed using the same fine-tuned embedding model employed for text document retrieval.
The multimodal data included in the multimodal answer are retrieved based on the cosine similarity between the text answer snippet $a_{[i, j]}$ and the text embeddings of multimodal data.
Only the highest-ranked multimodal data are selected for a text answer snippet. However, the same multimodal data may occasionally be selected for multiple text answer snippets, resulting in duplication within the multimodal answer. To address this, we only keep the multimodal data with the highest similarity score from the retrieval results.

\subsection{Multimodal Answer Refinement}
\label{sec:multimodal_answer_refinement}

After retrieving the multimodal data, an LLM is prompted to refine the initial text answer into a multimodal answer. The prompt includes the user question $q$, initial text answer $A_{t}$, and retrieved multimodal data accompanied by their contextual text features.

To guide the LLM to generate multimodal answers, placeholders are inserted into $A_{t}$ after the textual answer snippets corresponding to retrieved multimodal data.
Each placeholder includes the URL of the multimodal data and its contextual text features, ensuring the LLM incorporates relevant information while minimizing the risk of generating irrelevant details and hallucinations.
Additionally, the prompt includes several illustrative question-answering examples to demostrate how multimodal data should be integrated into the final answer. An example prompt is provided in Appendix~\ref{appendix:multimodal_prompt}. The LLM is instructed to replace the placeholders with appropriate descriptions and modify the text answer snippets to produce a coherent and readable multimodal response.
Finally, for display in the user interface, the URLs for multimodal data are converted into HTML elements, enabling users to interact with the multimodal content directly.

%% file: 04_user_interface.tex
\section{User Interface}

We implemented the \MuRAR framework and integrated it into a prototype version of the AI assistant within Adobe Experience Platform.\footnote{https://experienceleague.adobe.com/en/docs/experience-platform/ai-assistant/home} 
As illustrated in Figure~\ref{fig:UI}, when a user queries the AI assistant (A), for example by asking, "What is a good tutorial for creating a schema?", \MuRAR retrieves relevant text snippets (B) from documents on Adobe Experience League\footnote{https://experienceleague.adobe.com/en/docs} and generates an initial text answer.
Subsequently, source attribution is applied to each text answer snippet (C) to identify potential multimodal data candidates.
Next, \MuRAR retrieves the most relevant multimodal data (D1 and D2), which, in this case, include a screenshot of the Schemas Workspace and a video tutorial explaining the schema creation process. Finally, \MuRAR integrates the relevant multimodal data into the initial text answer through multimodal answer refinement.
Notably, the multimodal elements are interactive: when a user clicks on an element, a pop-up window appears, displaying the multimodal data in full size for detailed viewing.

%% file: 05_dataset.tex
\section{Data Collection}
\label{sec:dataset}

We curated two datasets to support the development of the \MuRAR framework and facilitate human evaluation. A multimodal document dataset was created to serve as $\mathcal{D}$ for developing the \MuRAR framework. Additionally, a multimodal question-answering dataset was collected and annotated for use in human evaluation.

\subsection{Multimodal Document Dataset}
The multimodal document dataset was collected from 2,173 web documents from Adobe Experience League~\footnote{\href{https://experienceleague.adobe.com/en/docs}{https://experienceleague.adobe.com/en/docs}}.
As summarized in Table~\ref{tab:dataset_statistics}, the dataset contains four modalities: text, image, table, and video. The textual data includes both plain text and tabular content, while the visual data comprises images and videos. Additional details about the data collection process can be found in Appendix~\ref{appendix:scraper}.

The text content was scraped from web documents and tokenized using the GPT-2 tokenizer~\cite{radford2019language}. Each document was segmented into smaller snippets ranging from 11 to 1,500 tokens, resulting in a total of 18,071 text snippets.
For image data, we extracted image URLs along with their surrounding textual context, which includes the text preceding the image (pre-image context) and the text following it (post-image context). To enrich the dataset with more text features for multimodal retrieval, GPT-4 was used to generate captions for the images, resulting in 6,320 annotated image entries.
Tables were extracted in JSON format, along with their associated contextual text. As with the image data, the contextual text for tables includes the text preceding and following the table. This process yielded 2,646 table entries.
For video content, we collected video URLs, contextual text, and transcripts. In cases where transcripts were unavailable, videos were downloaded, and Whisper~\cite{DBLP:conf/icml/RadfordKXBMS23} was used to generate transcripts from the audio. GPT-4 was then employed to summarize these transcripts, resulting in 253 video entries.
For all modalities, additional metadata were gathered, including the titles of the web document and the headings of the sections where the multimodal data were located.

\begin{table}[t]
    \centering
    \small
    \begin{tabular}{l l r}
        \toprule
        \textbf{Modality} & \textbf{Metric} & \textbf{Value} \\ \midrule
        \multirow{2}{*}{\textbf{Text}}
            & Count & 18,071 \\ \cline{2-3}
            & \rule{0pt}{2.5ex} Avg content tokens & 192 \\ \midrule
        \multirow{3}{*}{\textbf{Image}} 
            & Count & 6,320 \\ \cline{2-3}
            & \rule{0pt}{2.5ex}Avg context tokens & 238 \\ \cline{2-3}
            & \rule{0pt}{2.5ex}Avg caption tokens & 94 \\ \midrule
        \multirow{3}{*}{\textbf{Video}} 
            & Count & 253 \\ \cline{2-3}
            & \rule{0pt}{2.5ex}Avg context tokens & 91 \\ \cline{2-3}
            & \rule{0pt}{2.5ex}Avg summary tokens & 33 \\ \midrule
        \multirow{3}{*}{\textbf{Table}} 
            & Count & 2,644 \\ \cline{2-3}
            & \rule{0pt}{2.5ex}Avg context tokens & 160 \\ \cline{2-3}
            & \rule{0pt}{2.5ex}Avg table tokens & 223 \\ \bottomrule
    \end{tabular}
    \caption{Multimodal dataset statistics.}
    \label{tab:dataset_statistics}
\end{table}

\subsection{Multimodal Question Answering Dataset}
To construct a multimodal question-answer dataset, we collected 764 real customer questions and applied \MuRAR to generate answers. Among these, 306 questions were answered with multimodal data. For human evaluation, we randomly sampled 150 questions from this subset of 306 to assess the quality of the generated answers.
To analyze the impact of different backbone LLMs on the quality of multimodal answers, we conducted a comparative study using GPT-3.5 and GPT-4. For each model, we generated 150 text-based and multimodal answers, resulting in a total of 300 question-answer pairs. This evaluation enabled us to compare the effectiveness and coherence of multimodal responses across the two models, providing insights into how the choice of LLM influences performance.

%% file: 06_evaluation.tex
\section{Human Evaluation}
\label{sec:evaluation}

To assess the effectiveness of our multimodal question-answering system, we conducted a human evaluation study in two phases: (1) single-model evaluation and (2) pairwise comparison. The single-model psychometric evaluation was designed to measure three key metrics: usefulness, readability, and relevance of the multimodal answer. For the pairwise comparison, we used a preference-based ranking to determine the overall user preference between text-only and multimodal responses. Notably, we did not assess the quality of the text content itself but rather the added value of integrating multimodal data into the text answers.

\subsection{Study Setup}
We compiled a dataset of 300 question-answer pairs, evenly distributed between outputs from GPT-3.5 and GPT-4 models. For each question, both text-only and multimodal answers were generated. To evaluate the quality of these outputs, we recruited eight expert annotators, all holding advanced degrees in computer science with substantial experience in natural language processing (NLP). The annotation process was conducted on LabelStudio~\cite{LabelStudio}, with each question-answer pair evaluated by at least three experts.

\subsection{Evaluation Schema}

The annotators were asked to rate each multimodal answer on a 5-point Likert scale (1 being lowest, 5 being highest) for the following metrics, which were adapted from \cite{pradeep2024convkgyarnspinningconfigurablescalable}:

\vspace{-2mm}
\begin{itemize}[noitemsep,wide,topsep=10pt]
    \item \textbf{Usefulness:} This metric measures the extent to which multimodal elements contribute to the user's comprehension of the text content. High scores indicate that the multimodal output provides valuable additional information, clarifies complex concepts, or illustrates key points in ways that significantly aid understanding.
    \item \textbf{Readability:} This assesses how well the multimodal elements are integrated with the text, considering factors such as placement, size, and formatting. High scores indicate seamless integration that enhances the overall reading experience.
    \item \textbf{Relevance:} This measures how closely the multimodal elements relate to the content of the text. High scores indicate that the multimodal output directly supports or illustrates the textual content.
\end{itemize}

\begin{table}[ht]
\centering
\small
\begin{tabular}{l c c c}
\toprule
\multicolumn{1}{c}{\multirow{2}{*}{}} & \multicolumn{2}{c}{\textbf{Model}} \\ \cmidrule{2-3}
\textbf{Metric} & \textbf{GPT-3.5} & \textbf{GPT-4} & \textbf{Average} \\ \midrule
Usefulness & 3.34 & \textbf{3.60} & 3.47 \\ \midrule
Readability & 3.49 & \textbf{3.76} & 3.63 \\ \midrule
Relevance & 3.66 & \textbf{3.90} & 3.78 \\ \midrule
Preference Rate & 0.82 & \textbf{0.90} & 0.86 \\ \bottomrule
\end{tabular}
\caption{Evaluation results for multimodal answers generated by GPT-3.5 and GPT-4.}
\label{tab:evaluation_result}
\end{table}

\begin{figure*}[htbp]
    \centering
    \includegraphics[width=1\textwidth]{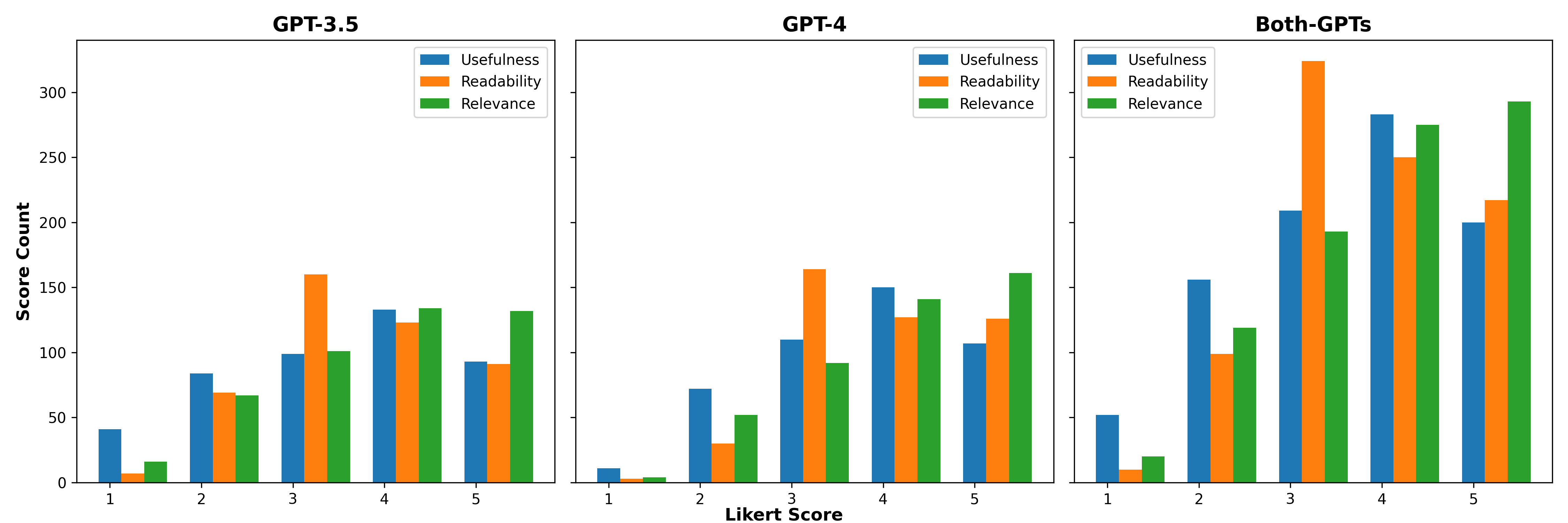}
    \caption{Distribution of score counts for GPT-3.5, GPT-4, and the combined results of GPT-3.5 and GPT-4.}
    \label{fig:evaluation_distribution}
\end{figure*}

After rating the multimodal answers, annotators were asked to indicate their overall preference between the text-only version, the multimodal version, or if they found them equally effective (``Same'').

\subsection{Results}
\paragraph{Psychometric Evaluation Results.}
The psychometric evaluation focused on three key aspects: usefulness, readability, and relevance. As shown in Table~\ref{tab:evaluation_result}, when examining the answers generated by both GPT-3.5 and GPT-4, the usefulness metric achieved an average score of 3.47, while readability and relevance scored 3.63 and 3.78, respectively. These scores, all above the midpoint of the scale as reflected in Figure \ref{fig:evaluation_distribution}, suggest that our approach performs well in producing useful, readable, and relevant output. Qualitative feedback from annotators further supports this conclusion, indicating that the multimodal answer provides informative additions to the text, is understandable through its placement, and remains relevant to the associated content. In addition, the average preference rating of 0.86 demonstrates a strong overall preference for our method compared to the text-only alternative. When comparing GPT-3.5 with GPT-4, we found that using GPT-4 as the backbone LLM increased all the metrics, showcasing its superior performance in generating high-quality, multimodal content within our framework.

\begin{table}[ht]
    \small
    \centering
    \begin{tabular}{lccc}
    \toprule
    \multicolumn{1}{c}{\multirow{2}{*}{}} & \multicolumn{3}{c}{\textbf{Agreement Metric}} \\ \cmidrule{2-4}
    \textbf{Metric\&Model} & \textbf{K-$\alpha$}\textsubscript{normal} & \textbf{K-$\alpha$}\textsubscript{combined} & \textbf{C-$\kappa$} \\
    \midrule
    Overall & 0.4179 & 0.3437 & 0.7100 \\
    \midrule
    Usefulness\textsubscript{GPT-3.5} & 0.4150 & 0.3468 & 0.6879 \\
    Usefulness\textsubscript{GPT-4} & 0.5424 & 0.4993 & 0.7900 \\
    Usefulness\textsubscript{all} & 0.4758 & 0.4164 & 0.7383 \\
    \midrule
    Readability\textsubscript{GPT-3.5} & 0.3418 & 0.2147 & 0.6852 \\
    Readability\textsubscript{GPT-4} & 0.3187 & 0.2502 & 0.7048 \\
    Readability\textsubscript{all} & 0.3424 & 0.2374 & 0.6968 \\
    \midrule
    Relevance\textsubscript{GPT-3.5} & 0.3369 & 0.2958 & 0.6459 \\
    Relevance\textsubscript{GPT-4} & 0.4465 & 0.3664 & 0.7291 \\
    Relevance\textsubscript{all} & 0.3925 & 0.3323 & 0.6872 \\
    \bottomrule
    \end{tabular}
    \caption{The inter-annotator agreement among the eight annotators.}
    \label{tab:iaa_table}
\end{table}

\begin{table}[ht]
    \centering
    \small
    \begin{tabular}{l c c c}
    \toprule
    \multicolumn{1}{c}{\multirow{2}{*}{}} & \multicolumn{2}{c}{\textbf{Model}} \\ \cmidrule{2-3}
    \textbf{Metric} & \textbf{GPT-3.5} & \textbf{GPT-4} & \textbf{GPT-3.5 \& GPT-4} \\ \midrule
    \textbf{Usefulness} & 0.9741 & 0.9059 & 0.9496 \\ \midrule
    \textbf{Readability} & 0.7686 & 0.7059 & 0.7500 \\ \midrule
    \textbf{Relevance} & 0.8576 & 0.8301 & 0.8519 \\ \bottomrule
    \end{tabular}
    \caption{Standard deviation of evaluation metrics for multimodal answers generated by GPT-3.5 and GPT-4.}
    \label{tab:standard_deviation_metrics}
\end{table}

\paragraph{Inter-Annotator Agreement.}
To assess annotation reliability, we calculated two inter-annotator agreement (IAA) measures (Table \ref{tab:iaa_table}). Krippendorff's alpha~\cite{Krippendorff2011ComputingKA} was 0.4179 overall, indicating moderate agreement across all annotators. However, Cohen's kappa~\cite{cohen1960coefficient} between the top two annotators was 0.71, suggesting substantial agreement when excluding outlier data points.
To investigate this discrepancy, we conducted further analyses. The standard deviation ranged from 0.8519 to 0.9496, indicating a relatively tight distribution of scores (Table \ref{tab:standard_deviation_metrics}). An annotator-specific analysis (Table \ref{tab:per_annotator}) revealed that Annotators 2 and 4, accounting for 26\% and 5.8\% of annotations, respectively, had lower average scores compared to others. Annotator 2 averaged 3.0619 (SD = 0.5648), while Annotator 4 averaged 3.3836 (SD = 0.253).
These findings suggest that the lower Krippendorff's alpha may be attributed to systematic differences in scoring patterns among a subset of annotators rather than widespread disagreement. It is worth noting that while annotators were provided with in-depth instructions, they did not undergo formal training. 

In conclusion, despite some variability in IAA results, the high Cohen's kappa for the top annotators, combined with strong psychometric evaluation scores and preference ratings, supports the overall reliability and effectiveness of our approach. The \MuRAR framwork demonstrates clear benefits over text-only alternatives, providing valuable enhancements to textual content.

\subsection{Limitations}
We analyzed the errors and mistakes made by the \MuRAR framework during our human evaluation. We identified some issues in the multimodal retrieval component. 
Although source attribution ensures precision, it can result in low recall, \textit{i.e.}, relevant multimodal data for a text answer snippet may not be in the same section or even the same web document. 
Additionally, readability can be affected by the multimodal answer refinement component. For instance, multimodal data may contain duplicated information already explained in plain text. This repetition can negatively impact the readability and clarity of the multimodal answer, making it less effective for users.

%% file: 07_conclusion.tex
\section{Conclusion and Future Work}

We introduced \MuRAR, a framework designed to enhance text-based responses by incorporating images, tables, and videos. Human evaluations showed that \MuRAR's multimodal answers are more useful, readable, and relevant than text-only responses. 
The system integrates text answer generation, source-based multimodal retrieval, and answer refinement to produce coherent multimodal answers.
Future work will focus on two key areas: improving the quality of multimodal answers and enhancing the user experience.
To improve the quality of multimodal answers, we plan to incorporate a broader range of multimodal documents to expand the dataset, train a custom LLM to replace reliance on proprietary models such as GPT-3.5, and address the relatively low recall performance in multimodal data retrieval.
To enhance the user experience, future developments could include features such as enabling videos to jump directly to time segments relevant to the user's question and providing a more intuitive and seamless interaction for users.

\section{Ethical Considerations}

\paragraph{Content Appropriateness.}
The \MuRAR framework integrates diverse data modalities, such as images and videos, into answers. 
The multimodal data used in this work is sourced from the Adobe Experience Platform, a business-focused document repository considered free from harmful content. However, our framework is designed for flexibility and can be applied to a wide range of multimodal question-answering systems and data sources.
This flexibility introduces the risk of retrieving inappropriate, offensive, or irrelevant material from poorly regulated or curated sources.
Such content could harm users or erode trust and safety.
To address this risk, it is essential to implement robust mechanisms for filtering, validating, and contextualizing multimodal content.

\paragraph{Privacy Concerns.}
Multimodal data retrieval could inadvertently expose sensitive information. Although the data source used in this work consists of enterprise-level documents devoid of sensitive content, the flexible nature of the \MuRAR framework allows integration with diverse data sources, where privacy concerns may arise. Privacy-preserving techniques are essential to protect sensitive user and organizational data during retrieval and integration, ensuring compliance with regulations and maintaining trust.

%% file: 08_appendix.tex
\section{Appendix}
\label{sec:appendix}

\begin{table*}[ht]
    \small
    \centering
    \begin{tabular}{lcccccc}
    \toprule
    \textbf{Annotator} & \textbf{Answers} & \textbf{Usefulness} & \textbf{Readability} & \textbf{Relevance} & \textbf{Preference (Multi-Modal / Text Only / Same)} \\
    \midrule
    No.1 & 294 & 3.6463 & 3.8367 & 4.0170 & 207 (70.41\%) / 41 (13.95\%) / 46 (15.65\%) \\
    \midrule
    No.2 & 237 & 2.8861 & 3.0295 & 3.2700 & 115 (48.52\%) / 81 (34.18\%) / 41 (17.30\%) \\
    \midrule
    No.3 & 259 & 3.7452 & 3.8340 & 3.8764 & 205 (79.15\%) / 28 (10.81\%) / 26 (10.04\%) \\
    \midrule
    No.4 & 53 & 3.0566 & 3.5094 & 3.5849 & 30 (56.60\%) / 16 (30.19\%) / 7 (13.21\%) \\
    \midrule
    No.5 & 9 & 4.1111 & 4.4444 & 4.4444 & 9 (100.0\%) / 0 (0.0\%) / 0 (0.0\%) \\
    \midrule
    No.6 & 22 & 4.0000 & 4.0909 & 4.1818 & 19 (86.36\%) / 2 (9.09\%) / 1 (4.55\%) \\
    \midrule
    No.7 & 24 & 4.2500 & 4.2083 & 4.6250 & 23 (95.83\%) / 0 (0.0\%) / 0 (0.0\%) \\
    \midrule
    No.8 & 2 & 4.0000 & 4.5000 & 4.5000 & 2 (100.0\%) / 0 (0.0\%) / 0 (0.0\%) \\
    \bottomrule
    \end{tabular}
    \caption{Per-annotator average scores and preference.}
    \vspace{-5mm}
    \label{tab:per_annotator}
\end{table*}

\subsection{Multimodal Scraper Design}
\label{appendix:scraper}
Our multimodal scraper design collects various fields and metadata from Adobe Experience League for images, videos, and tables. For images, we gather the link to the image and the surrounding context, specifically the text between the previous and current image and between the current and next image. The metadata collected includes the title of the document, the header of each section containing the image, and the URL of the document. For videos, the fields collected include the URL of the video, the context text before the video, and the video transcript. The metadata gathered is similar, including the document title, section headers, and document URL. For tables, we collect the table content in the form of a JSON string, the context text before the table, and the document URL. Additional metadata includes the document title and the header of each section containing the table.

\subsection{Human Evaluation Metrics}
\label{appendix:evaluation_metrics}
\paragraph{Usefulness}
Usefulness measures how much the multimodal elements contribute to the user's comprehension of the text content.
\begin{itemize}[noitemsep,wide,topsep=1pt]
\item 1 - Not at all useful: Multimodal elements provide no additional understanding or actively confuse the user.
\item 2 - Slightly useful: Multimodal elements offer minimal enhancement to understanding.
\item 3 - Moderately useful: Multimodal elements provide some additional clarity or information.
\item 4 - Very useful: Multimodal elements significantly enhance understanding of the text.
\item 5 - Extremely useful: Multimodal elements are crucial for full comprehension of the text.
\end{itemize}

\paragraph{Readability}
Readability assesses how well the multimodal elements are integrated with the text.
\begin{itemize}[noitemsep,wide,topsep=1pt]
\item 1 - Severely impairs readability: Multimodal elements are poorly placed, causing significant disruption to reading flow.
\item 2 - Somewhat impairs readability: Multimodal elements are not well-integrated, causing minor disruptions.
\item 3 - Neutral impact on readability: Multimodal elements neither enhance nor impair the reading experience.
\item 4 - Enhances readability: Multimodal elements are well-placed, supporting smooth reading flow.
\item 5 - Significantly enhances readability: Multimodal elements are perfectly integrated, greatly improving the reading experience.
\end{itemize}

\paragraph{Relevance}
Relevance measures how closely the multimodal elements relate to the text content.
\begin{itemize}[noitemsep,wide,topsep=1pt]
\item 1 - Completely irrelevant: Multimodal elements have no apparent connection to the text.
\item 2 - Mostly irrelevant: Multimodal elements have only a tenuous connection to the text.
\item 3 - Somewhat relevant: Multimodal elements relate to the text but not be entirely on-point.
\item 4 - Highly relevant: Multimodal elements clearly support and illustrate the text content.
\item 5 - Perfectly relevant: Multimodal elements are essential to the text, providing crucial illustrations or data.
\end{itemize}

\paragraph{Preference Grading}

Annotators also indicate their overall preference between the text answer and the multimodal answer:
\begin{itemize}[noitemsep,wide,topsep=1pt]
\item \textbf{Text Only}: Choose this if you believe the text alone would be more effective without the multimodal elements.
\item \textbf{Multi-Modal}: Select this if you think the combination of text and multimodal elements provides the best experience.
\item \textbf{Same}: Choose this if you feel text-only and multimodal versions are equally effective.
\end{itemize}

\subsection{Additional Human Evaluation Results}
\label{appendix:}

Due to space constraints, we include additional human evaluation results in the Appendix. The average score and preference per annotator are presented in Table~\ref{tab:per_annotator}.

\subsection{Prompts}
Please note that the actual prompts used in the system development differ from the prompts shown below. These are simplified versions that capture the essence of the prompt design.

\begin{figure}[ht]
    \centering
    \includegraphics[width=0.48\textwidth]{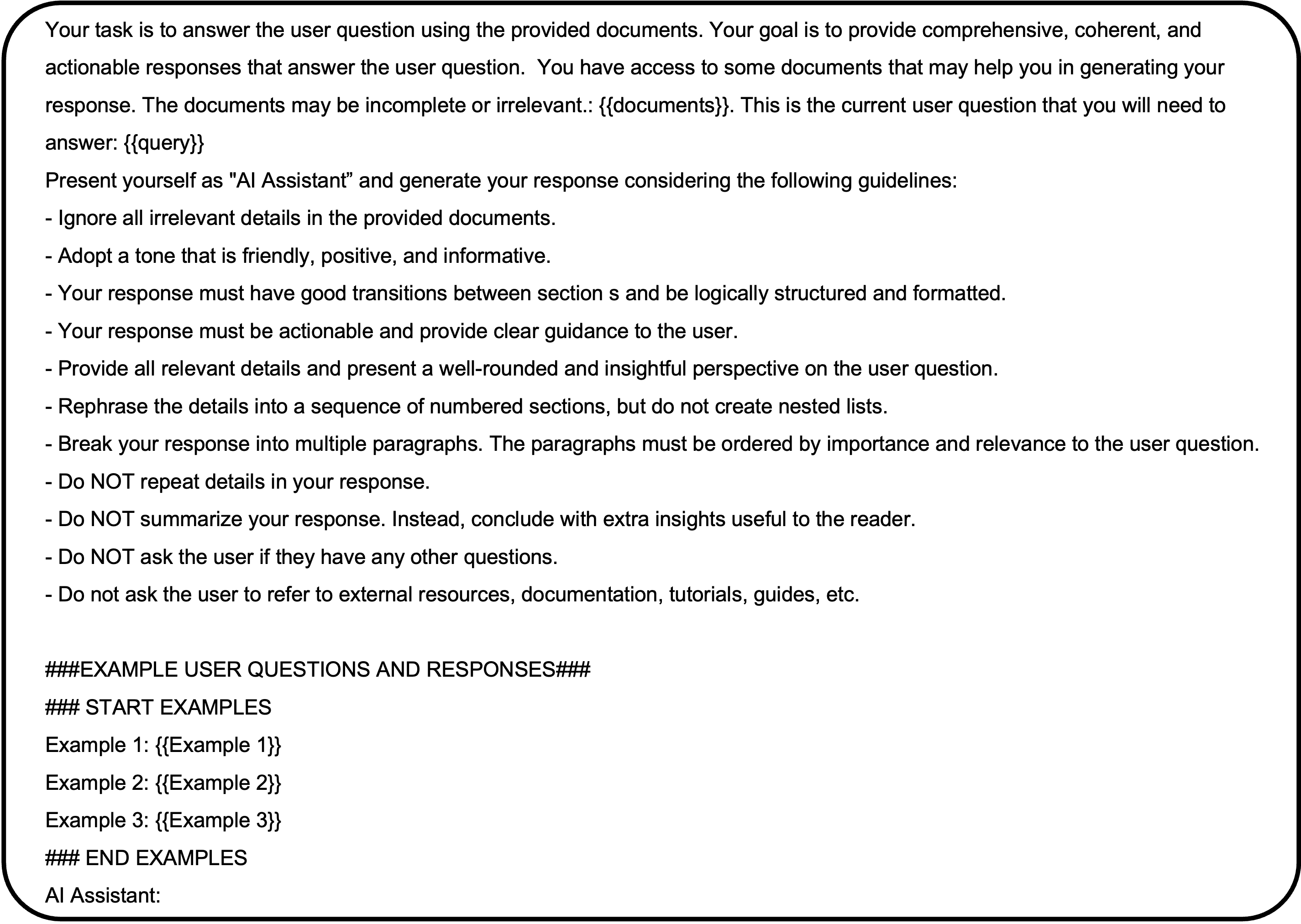}
    \caption{Prompt for text answer generation.}
    \vspace{-6mm}
    \label{fig:text_answer_prompt}
\end{figure}

\paragraph{Prompt for Text Answer Generation}
\label{appendix:text_answer_prompt}
The prompt for text answer generation can be found in Figure~\ref{fig:text_answer_prompt}.

\begin{figure}[ht]
    \centering
    \includegraphics[width=0.48\textwidth]{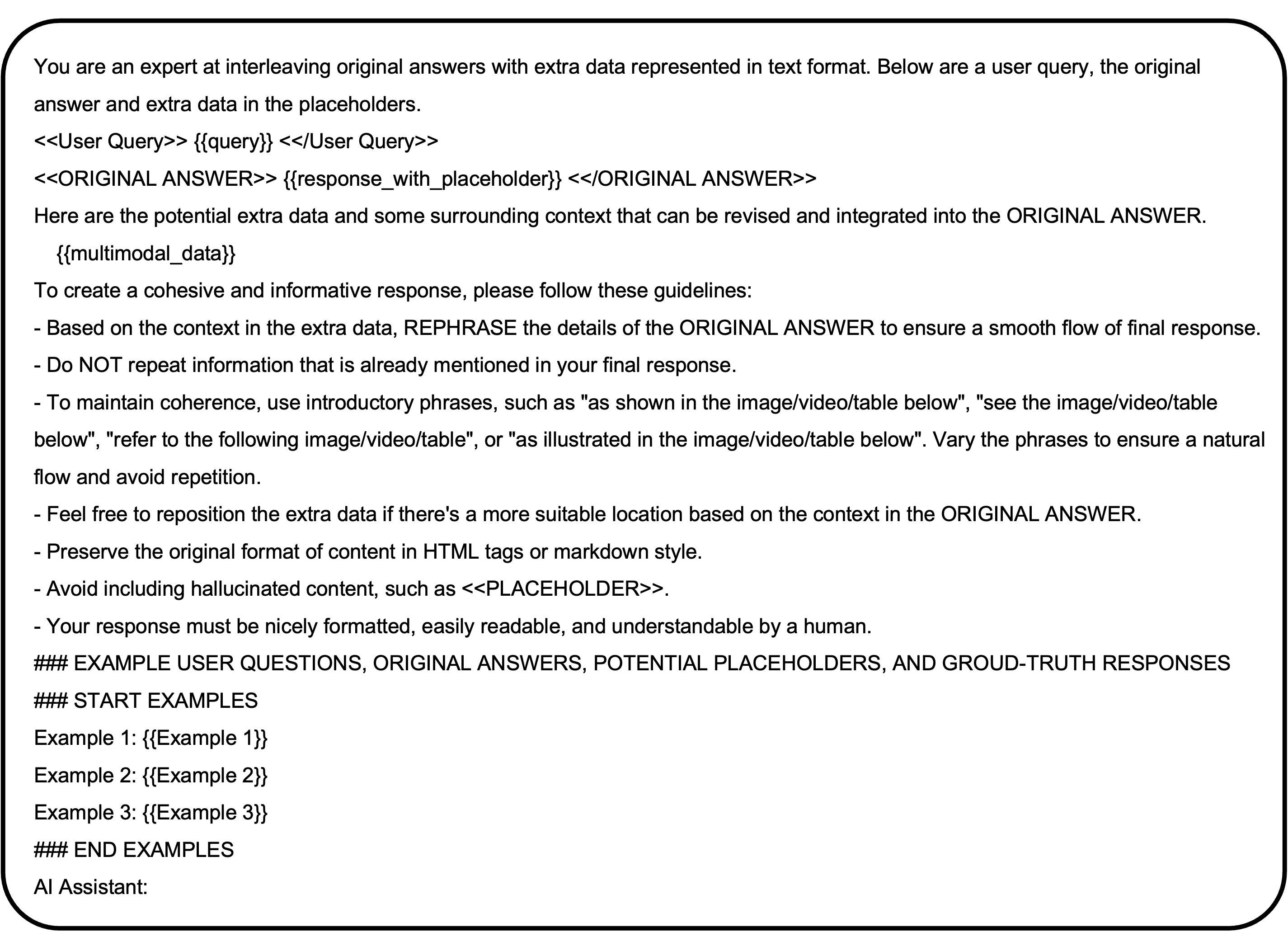}
    \caption{Prompt for multimodal answer refinement.}
    \label{fig:multimodal_answer_prompt}
\end{figure}

\paragraph{Prompt for Multimodal Answer Refinement}
\label{appendix:multimodal_prompt}
The prompt for multimodal answer refinement can be found in Figure~\ref{fig:multimodal_answer_prompt}. 

%% file: emnlp2023.bbl
\begin{thebibliography}{21}
\expandafter\ifx\csname natexlab\endcsname\relax\def\natexlab#1{#1}\fi

\bibitem[{Anil et~al.(2023)Anil, Borgeaud, Wu, Alayrac, Yu, Soricut, Schalkwyk, Dai, Hauth, Millican, Silver, Petrov, Johnson, Antonoglou, Schrittwieser, Glaese, Chen, Pitler, Lillicrap, Lazaridou, Firat, Molloy, Isard, Barham, Hennigan, Lee, Viola, Reynolds, Xu, Doherty, Collins, Meyer, Rutherford, Moreira, Ayoub, Goel, Tucker, Piqueras, Krikun, Barr, Savinov, Danihelka, Roelofs, White, Andreassen, von Glehn, Yagati, Kazemi, Gonzalez, Khalman, Sygnowski, and et~al.}]{DBLP:journals/corr/abs-2312-11805}
Rohan Anil, Sebastian Borgeaud, Yonghui Wu, Jean{-}Baptiste Alayrac, Jiahui Yu, Radu Soricut, Johan Schalkwyk, Andrew~M. Dai, Anja Hauth, Katie Millican, David Silver, Slav Petrov, Melvin Johnson, Ioannis Antonoglou, Julian Schrittwieser, Amelia Glaese, Jilin Chen, Emily Pitler, Timothy~P. Lillicrap, Angeliki Lazaridou, Orhan Firat, James Molloy, Michael Isard, Paul~Ronald Barham, Tom Hennigan, Benjamin Lee, Fabio Viola, Malcolm Reynolds, Yuanzhong Xu, Ryan Doherty, Eli Collins, Clemens Meyer, Eliza Rutherford, Erica Moreira, Kareem Ayoub, Megha Goel, George Tucker, Enrique Piqueras, Maxim Krikun, Iain Barr, Nikolay Savinov, Ivo Danihelka, Becca Roelofs, Ana{\"{\i}}s White, Anders Andreassen, Tamara von Glehn, Lakshman Yagati, Mehran Kazemi, Lucas Gonzalez, Misha Khalman, Jakub Sygnowski, and et~al. 2023.
\newblock \href {https://doi.org/10.48550/ARXIV.2312.11805} {Gemini: {A} family of highly capable multimodal models}.
\newblock \emph{CoRR}, abs/2312.11805.

\bibitem[{Brown et~al.(2020)Brown, Mann, Ryder, Subbiah, Kaplan, Dhariwal, Neelakantan, Shyam, Sastry, Askell, Agarwal, Herbert{-}Voss, Krueger, Henighan, Child, Ramesh, Ziegler, Wu, Winter, Hesse, Chen, Sigler, Litwin, Gray, Chess, Clark, Berner, McCandlish, Radford, Sutskever, and Amodei}]{DBLP:conf/nips/BrownMRSKDNSSAA20}
Tom~B. Brown, Benjamin Mann, Nick Ryder, Melanie Subbiah, Jared Kaplan, Prafulla Dhariwal, Arvind Neelakantan, Pranav Shyam, Girish Sastry, Amanda Askell, Sandhini Agarwal, Ariel Herbert{-}Voss, Gretchen Krueger, Tom Henighan, Rewon Child, Aditya Ramesh, Daniel~M. Ziegler, Jeffrey Wu, Clemens Winter, Christopher Hesse, Mark Chen, Eric Sigler, Mateusz Litwin, Scott Gray, Benjamin Chess, Jack Clark, Christopher Berner, Sam McCandlish, Alec Radford, Ilya Sutskever, and Dario Amodei. 2020.
\newblock \href {https://proceedings.neurips.cc/paper/2020/hash/1457c0d6bfcb4967418bfb8ac142f64a-Abstract.html} {Language models are few-shot learners}.
\newblock In \emph{Advances in Neural Information Processing Systems 33: Annual Conference on Neural Information Processing Systems 2020, NeurIPS 2020, December 6-12, 2020, virtual}.

\bibitem[{Cohen(1960)}]{cohen1960coefficient}
Jacob Cohen. 1960.
\newblock A coefficient of agreement for nominal scales.
\newblock \emph{Educational and psychological measurement}, 20(1):37--46.

\bibitem[{Gao et~al.(2023)Gao, Xiong, Gao, Jia, Pan, Bi, Dai, Sun, Guo, Wang, and Wang}]{DBLP:journals/corr/abs-2312-10997}
Yunfan Gao, Yun Xiong, Xinyu Gao, Kangxiang Jia, Jinliu Pan, Yuxi Bi, Yi~Dai, Jiawei Sun, Qianyu Guo, Meng Wang, and Haofen Wang. 2023.
\newblock \href {https://doi.org/10.48550/ARXIV.2312.10997} {Retrieval-augmented generation for large language models: {A} survey}.
\newblock \emph{CoRR}, abs/2312.10997.

\bibitem[{Johnson et~al.(2019)Johnson, Douze, and J{\'e}gou}]{johnson2019billion}
Jeff Johnson, Matthijs Douze, and Herv{\'e} J{\'e}gou. 2019.
\newblock Billion-scale similarity search with {GPUs}.
\newblock \emph{IEEE Transactions on Big Data}, 7(3):535--547.

\bibitem[{Joshi et~al.(2024)Joshi, Gupta, Kumar, and Sisodia}]{joshi2024robust}
Pankaj Joshi, Aditya Gupta, Pankaj Kumar, and Manas Sisodia. 2024.
\newblock Robust multi model rag pipeline for documents containing text, table \& images.
\newblock In \emph{2024 3rd International Conference on Applied Artificial Intelligence and Computing (ICAAIC)}, pages 993--999. IEEE.

\bibitem[{Krippendorff(2011)}]{Krippendorff2011ComputingKA}
Klaus Krippendorff. 2011.
\newblock \href {https://api.semanticscholar.org/CorpusID:59901023} {Computing krippendorff's alpha-reliability}.

\bibitem[{Kumar et~al.(2020)Kumar, Mittal, and Manocha}]{DBLP:journals/corr/abs-2004-12238}
Abhishek Kumar, Trisha Mittal, and Dinesh Manocha. 2020.
\newblock \href {http://arxiv.org/abs/2004.12238} {{MCQA:} multimodal co-attention based network for question answering}.
\newblock \emph{CoRR}, abs/2004.12238.

\bibitem[{Lewis et~al.(2020)Lewis, Perez, Piktus, Petroni, Karpukhin, Goyal, K{\"{u}}ttler, Lewis, Yih, Rockt{\"{a}}schel, Riedel, and Kiela}]{DBLP:conf/nips/LewisPPPKGKLYR020}
Patrick S.~H. Lewis, Ethan Perez, Aleksandra Piktus, Fabio Petroni, Vladimir Karpukhin, Naman Goyal, Heinrich K{\"{u}}ttler, Mike Lewis, Wen{-}tau Yih, Tim Rockt{\"{a}}schel, Sebastian Riedel, and Douwe Kiela. 2020.
\newblock \href {https://proceedings.neurips.cc/paper/2020/hash/6b493230205f780e1bc26945df7481e5-Abstract.html} {Retrieval-augmented generation for knowledge-intensive {NLP} tasks}.
\newblock In \emph{Advances in Neural Information Processing Systems 33: Annual Conference on Neural Information Processing Systems 2020, NeurIPS 2020, December 6-12, 2020, virtual}.

\bibitem[{OpenAI(2023)}]{DBLP:journals/corr/abs-2303-08774}
OpenAI. 2023.
\newblock \href {https://doi.org/10.48550/ARXIV.2303.08774} {{GPT-4} technical report}.
\newblock \emph{CoRR}, abs/2303.08774.

\bibitem[{Pradeep et~al.(2024)Pradeep, Lee, Mousavi, Pound, Sang, Lin, Ilyas, Potdar, Arefiyan, and Li}]{pradeep2024convkgyarnspinningconfigurablescalable}
Ronak Pradeep, Daniel Lee, Ali Mousavi, Jeff Pound, Yisi Sang, Jimmy Lin, Ihab Ilyas, Saloni Potdar, Mostafa Arefiyan, and Yunyao Li. 2024.
\newblock \href {http://arxiv.org/abs/2408.05948} {Convkgyarn: Spinning configurable and scalable conversational knowledge graph qa datasets with large language models}.

\bibitem[{Radford et~al.(2023)Radford, Kim, Xu, Brockman, McLeavey, and Sutskever}]{DBLP:conf/icml/RadfordKXBMS23}
Alec Radford, Jong~Wook Kim, Tao Xu, Greg Brockman, Christine McLeavey, and Ilya Sutskever. 2023.
\newblock \href {https://proceedings.mlr.press/v202/radford23a.html} {Robust speech recognition via large-scale weak supervision}.
\newblock In \emph{International Conference on Machine Learning, {ICML} 2023, 23-29 July 2023, Honolulu, Hawaii, {USA}}, volume 202 of \emph{Proceedings of Machine Learning Research}, pages 28492--28518. {PMLR}.

\bibitem[{Radford et~al.(2019)Radford, Wu, Child, Luan, Amodei, Sutskever et~al.}]{radford2019language}
Alec Radford, Jeffrey Wu, Rewon Child, David Luan, Dario Amodei, Ilya Sutskever, et~al. 2019.
\newblock Language models are unsupervised multitask learners.
\newblock \emph{OpenAI blog}, 1(8):9.

\bibitem[{Reimers and Gurevych(2019)}]{reimers-2019-sentence-bert}
Nils Reimers and Iryna Gurevych. 2019.
\newblock \href {https://arxiv.org/abs/1908.10084} {Sentence-bert: Sentence embeddings using siamese bert-networks}.
\newblock In \emph{Proceedings of the 2019 Conference on Empirical Methods in Natural Language Processing}. Association for Computational Linguistics.

\bibitem[{Singh et~al.(2021)Singh, Nasery, Mehta, Agarwal, Lamba, and Srinivasan}]{DBLP:conf/naacl/SinghNMALS21}
Hrituraj Singh, Anshul Nasery, Denil Mehta, Aishwarya Agarwal, Jatin Lamba, and Balaji~Vasan Srinivasan. 2021.
\newblock \href {https://doi.org/10.18653/V1/2021.NAACL-MAIN.418} {{MIMOQA:} multimodal input multimodal output question answering}.
\newblock In \emph{Proceedings of the 2021 Conference of the North American Chapter of the Association for Computational Linguistics: Human Language Technologies, {NAACL-HLT} 2021, Online, June 6-11, 2021}, pages 5317--5332. Association for Computational Linguistics.

\bibitem[{Talmor et~al.(2021)Talmor, Yoran, Catav, Lahav, Wang, Asai, Ilharco, Hajishirzi, and Berant}]{DBLP:conf/iclr/TalmorYCLWAIHB21}
Alon Talmor, Ori Yoran, Amnon Catav, Dan Lahav, Yizhong Wang, Akari Asai, Gabriel Ilharco, Hannaneh Hajishirzi, and Jonathan Berant. 2021.
\newblock \href {https://openreview.net/forum?id=ee6W5UgQLa} {Multimodalqa: complex question answering over text, tables and images}.
\newblock In \emph{9th International Conference on Learning Representations, {ICLR} 2021, Virtual Event, Austria, May 3-7, 2021}. OpenReview.net.

\bibitem[{Tkachenko et~al.(2020-2022)Tkachenko, Malyuk, Holmanyuk, and Liubimov}]{LabelStudio}
Maxim Tkachenko, Mikhail Malyuk, Andrey Holmanyuk, and Nikolai Liubimov. 2020-2022.
\newblock \href {https://github.com/heartexlabs/label-studio} {{Label Studio}: Data labeling software}.
\newblock Open source software available from https://github.com/heartexlabs/label-studio.

\bibitem[{Touvron et~al.(2023)Touvron, Lavril, Izacard, Martinet, Lachaux, Lacroix, Rozi{\`{e}}re, Goyal, Hambro, Azhar, Rodriguez, Joulin, Grave, and Lample}]{DBLP:journals/corr/abs-2302-13971}
Hugo Touvron, Thibaut Lavril, Gautier Izacard, Xavier Martinet, Marie{-}Anne Lachaux, Timoth{\'{e}}e Lacroix, Baptiste Rozi{\`{e}}re, Naman Goyal, Eric Hambro, Faisal Azhar, Aur{\'{e}}lien Rodriguez, Armand Joulin, Edouard Grave, and Guillaume Lample. 2023.
\newblock \href {https://doi.org/10.48550/ARXIV.2302.13971} {Llama: Open and efficient foundation language models}.
\newblock \emph{CoRR}, abs/2302.13971.

\bibitem[{Wei et~al.(2022)Wei, Wang, Schuurmans, Bosma, Ichter, Xia, Chi, Le, and Zhou}]{DBLP:conf/nips/Wei0SBIXCLZ22}
Jason Wei, Xuezhi Wang, Dale Schuurmans, Maarten Bosma, Brian Ichter, Fei Xia, Ed~H. Chi, Quoc~V. Le, and Denny Zhou. 2022.
\newblock \href {http://papers.nips.cc/paper\_files/paper/2022/hash/9d5609613524ecf4f15af0f7b31abca4-Abstract-Conference.html} {Chain-of-thought prompting elicits reasoning in large language models}.
\newblock In \emph{Advances in Neural Information Processing Systems 35: Annual Conference on Neural Information Processing Systems 2022, NeurIPS 2022, New Orleans, LA, USA, November 28 - December 9, 2022}.

\bibitem[{Wu et~al.(2023)Wu, Fei, Qu, Ji, and Chua}]{DBLP:journals/corr/abs-2309-05519}
Shengqiong Wu, Hao Fei, Leigang Qu, Wei Ji, and Tat{-}Seng Chua. 2023.
\newblock \href {https://doi.org/10.48550/ARXIV.2309.05519} {Next-gpt: Any-to-any multimodal {LLM}}.
\newblock \emph{CoRR}, abs/2309.05519.

\bibitem[{Zhang et~al.(2024)Zhang, Yu, Zhang, Li, Zhong, Liang, Yan, Ma, Weng, Pan, Li, Xu, and Lan}]{DBLP:journals/corr/abs-2406-15000}
Lichao Zhang, Jia Yu, Shuai Zhang, Long Li, Yangyang Zhong, Guanbao Liang, Yuming Yan, Qing Ma, Fangsheng Weng, Fayu Pan, Jing Li, Renjun Xu, and Zhenzhong Lan. 2024.
\newblock \href {https://doi.org/10.48550/ARXIV.2406.15000} {Unveiling the impact of multi-modal interactions on user engagement: {A} comprehensive evaluation in ai-driven conversations}.
\newblock \emph{CoRR}, abs/2406.15000.

\end{thebibliography}
